\documentclass[prl,showpacs,twocolumn,aps]{revtex4}
\usepackage{amsmath}
\usepackage{graphicx}
\usepackage{epsfig}
\usepackage{amsfonts}
\usepackage{amssymb}
\begin{document}
\title{ Simple Algorithm for Partial Quantum Search}
\author{Vladimir\ E.\ Korepin$^{1}$ and Lov K.\ Grover$^{2}$ }
\affiliation{$^{1}$C.N.\ Yang Institute for Theoretical Physics, State University of New
York at Stony Brook,}
\affiliation{Stony Brook, NY 11794-3840 e-mail: korepin@insti.physics.sunysb.edu }
\affiliation{$^{2}$Bell Laboratories, Lucent Technologies, 600--700 Mountain
Avenue, Murray Hill, NJ 07974, e-mail: lkgrover@bell-labs.com}
\date{\today}

\begin{abstract}
Quite often in database search, we only need to extract portion of the
information about the satisfying item. Recently Radhakrishnan and Grover [RG]
considered this problem in the
following form: the database of $N$ items was divided into $K$
equally sized blocks. The algorithm has just to find the block
containing the item of interest. The queries are exactly the same as in the
standard database search problem. [RG] invented a quantum algorithm for this
problem of partial search that  took about $0.33\sqrt{N/K}$ fewer iterations
than the quantum search algorithm. They also proved that the best any
quantum algorithm could do
would be to save $0.78 \sqrt(N/K)$ iterations. The main limitation of
the algorithm
was that it involved complicated analysis as a result of which it has
been inaccessible
to most of the community. This paper gives a simple analysis of the
algorithm. This
analysis is based on three elementary observations about quantum search,
does not require
a single equation and takes less than 2 pages.

\end{abstract}

\pacs{03.67.-a, 03.67.Lx}
\maketitle


Database search is one of the few applications for which a fast quantum
algorithm is known
\cite{Grover}. The Grover algorithm has been shown to be optimal and cannot be
improved even by a single query \cite{Zalka, NC}. Therefore it is of great
interest to find any circumstance under which we can improve the performance
of the quantum search algorithm. The problem of partial database search is
also of independent interest
. For example when using Google to search the internet (which is a large
database), we are typically interested in only some of the attributes of the
entity being searched - e.g. while searching for a grocery store in the
neighborhood, we may only want the address of the store, not its corporate
information.
\begin{figure}[ht]
\begin{center}
\epsfig{figure=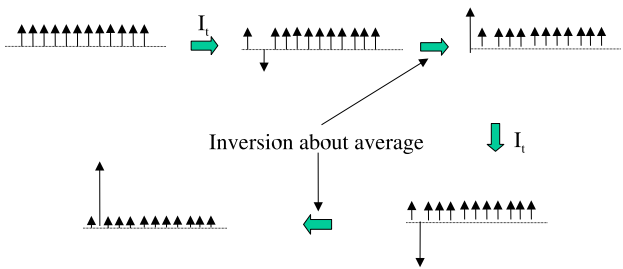, scale=0.35}
\caption{Alternately repeating the two operations shown above,  drives
amplitude into the target state.  }
\end{center}
\end{figure}

The concept of partial search was invented recently in \cite{jaik}. That was a
surprising result because it gave a means for improving the quantum search
algorithm with no knowledge of the problem structure, except of course the
block nature. Unfortunately, that paper, though an important discovery, is
mathematically, very rigorous and not accessible to most of the quantum
information community. In this paper we set ourselves the task of designing a
simple partial search algorithm that clearly brought out the nature of the
algorithm. 

The original Grover search algorithm is based on two operations: (i)
selective inversion and (ii) inversion about average. We shall denote a
selective inversion of the target, followed by an inversion about average, as
a Grover iteration. Figure 1 depicts this sequence of two operations which is
being used to search for a single target out of twelve items. In the original
algorithm, it takes $({\pi}/{4})\sqrt{N}$ Grover iterations to locate the
desired (target) item.
The idea of partial search is a trade-off of precision for speed, i.e. we do
not need the exact address of the target, but only the first several bits of
it as illustrated in figure 2.
\begin{figure}[ht]
\begin{center}
\epsfig{figure=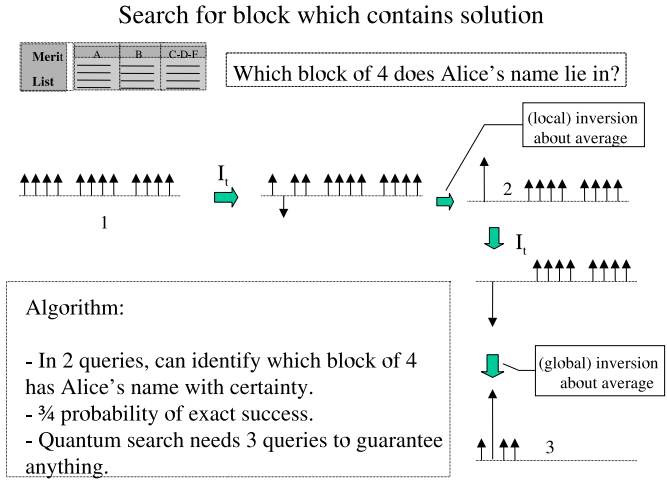, scale=0.35}
\caption{ A partial quantum search is able to find partial information about
the solution faster than the complete quantum search can. }
\end{center}
\end{figure}


Let us consider a more general setup \textbf{$N$ items are divided into $K$
blocks of $b=N/K$ items each}. As in \cite{jaik}, we do a partial search for
the appropriate block.  The following are the three conceptual steps of the
algorithm:
\begin{figure}[ht]
\begin{center}
\epsfig{figure=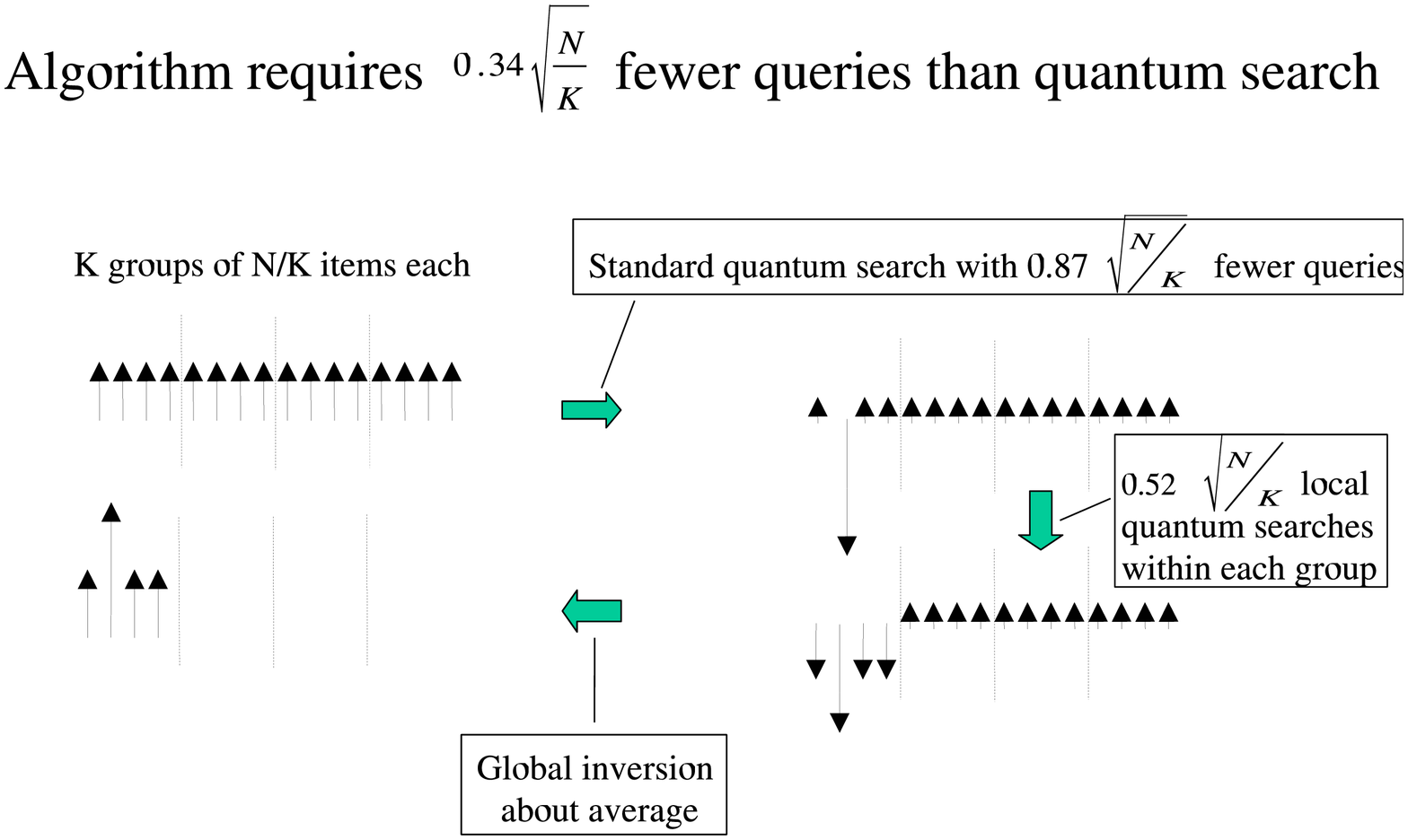, scale=0.35}
\caption{ The three arrows correspond to the three steps of the partial quantum
search algorithm. }
\end{center}
\end{figure}


\begin{itemize}
\item Step 1: $\frac{\pi}{4}\sqrt{N}-\sqrt{\frac{3b}{4}}$ Grover iterations
(global search).

\item Step 2: $\frac{\pi}{6}\sqrt{b}$ iterations of local searches in each
block done in parallel. Note that this drives the amplitude negative in the
target block.

\item Step 3: one global inversion about average annihilates amplitudes of all
items in non-target blocks and finds the target block
\end{itemize}

So the total number of queries $\left(  Q\right)  $ is
\[
Q=\frac{\pi}{4}\sqrt{N}-\left(  \sqrt{\frac{3}{4}}-\frac{\pi}{6}\right)
\sqrt{b}.
\]
The coefficient of $\sqrt{b}$ is
$(\sqrt{3/4}-\pi/6)\approx0.34$ and so the improvement over quantum searching
is $0.34\sqrt{b}$ iterations, just for reference the original algorithm of
\cite{jaik} achieved a saving of $0.33\sqrt{b}$ iterations after considerably
more work.

In order to see qualitatively, why the above procedure works and why this is
the best we can do within this framework, consider the following argument
based on three simple observations ((a), (b)\ \& (c)):

\begin{description}
\item[a. Scattering out of a state -] If we start with all the amplitude in a
single basis state, and apply $\eta$ iterations of Grover search, then the sum of
the amplitudes in all $b$ states will be $\sqrt{b}\sin\frac{2\eta}{\sqrt{b}}$.

\item[b. Going into a state] - If we stop the search algorithm $\eta$
iterations before it finds the target, then the sum of amplitudes in all
$N$\ states will be $\sqrt{N}\sin\frac{2\eta}{\sqrt{N}}$.$\ $In case
$\eta<<\sqrt{N},$ this sum becomes $2\eta$.

\item[c. \ Zeroing the amplitudes in certain states] - If the amplitude in
some state has to fall to zero after a Grover iteration (step 3 of our
algorithm), the state should have an amplitude of two times the average before
the iteration$.$
\end{description}

Working backward from the final result:

\begin{itemize}
\item Assume $\eta$ iterations in step 2, and neglecting the initial
iterations it takes for the amplitudes in the target block to come to zero, it
follows by (a), that after step 2, the sum of the amplitudes in the target
block should be $-\sqrt{b}\sin\frac{2\eta}{\sqrt{b}}$.

\item Therefore in order for step 3 to work, it follows by (c), the sum of the
amplitude in all states (in non-target blocks), after step 1, should have been
$2\sqrt{b}\sin\frac{2\eta}{\sqrt{b}}.$

\item Therefore by (b), the saving in step 1 \ is $\frac{1}{2}\times2\sqrt
{b}\sin\frac{2\eta}{\sqrt{b}}$ Grover iterations$.$

\item This gives an overall saving of $-\eta+\sqrt{b}\sin\frac{2\eta}{\sqrt
{b}}$ iterations. This function assumes its maximum value at $\eta=\frac{\pi
}{6}\sqrt{b}$ which is $\sqrt{b}\left(  -\frac{\pi}{6}+\frac{\sqrt{3}}%
{2}\right)  .$
\end{itemize}

\section{Lower Bound}

It is relatively easy to find a lower bound for the total number of queries,
$S$.
Let us try to find the target by first locating the target block and then use
Grover's algorithm to find the target in the block. We know that it takes
$(\pi/4)\sqrt{b}$ queries to find the target in a block. The overall number of
queries should be greater than $({\pi}/{4})\sqrt{N}$ \cite{Zalka}.
Therefore $Q+({\pi}/{4})\sqrt{b}\geq({\pi}/{4})\sqrt{N}$
. This gives a lower bound for $Q$ : $\ Q\geq ({\pi}/{4})\sqrt{N}-({\pi}/{4})\sqrt{b}.$The lower bound is rather obvious, the existence of an
algorithm, that was discovered in \cite{jaik} that came so close to the lower
bound was a lot more surprising.

\section{Summary}

In the paper we have presented a simple algorithm for partial search of a
database of $N$ items separated into $K$ blocks of $b$ items each, $N=Kb$. The
saving in the run-time as compared to an exhaustive search, is slightly better
than the original partial quantum search algorithm. However, the
distinguishing feature is not the savings but its simplicity. Through three
elementary observations about the Grover search algorithm, without a single
equation, we derive one of the fastest possible search algorithms.

\section{Acknowledgments}

Research was partly supported by NSA , ARO under contract no.~DAAG55-98-C-0040.



\end{document}